\documentclass[11pt]{article}
\usepackage{moriond,epsfig}

\def\be{\begin{equation}}
\def\ee{\end{equation}}
\def\bea{\begin{eqnarray}}
\def\eea{\end{eqnarray}}

\begin{document}
\vspace*{4cm}
\title{DISSIPATION, COLLECTIVE FLOW AND MACH CONES AT RHIC}

\author{I. BOURAS, A. EL, O. FOCHLER, J. UPHOFF, Z. XU AND C. GREINER  }

\address{Institut f\"ur Theoretische Physik, Goethe-Universit\"at Frankfurt,  Max-von-Laue-Str.~1,\\ D-60438 Frankfurt am Main, Germany}

\maketitle

\abstracts{Fast thermalization and a 
strong buildup of elliptic flow of QCD matter 
as found at RHIC are understood as the consequence
of perturbative QCD (pQCD) interactions within the 3+1 dimensional parton
cascade BAMPS. The main contributions stem from pQCD bremsstrahlung 
$2 \leftrightarrow 3 $ processes.
By comparing to Au+Au data of the flow parameter $v_2$
the shear viscosity to entropy ratio $\eta/s$ has been extracted 
dynamically and lies in the range of
$0.08$ and $0.2$. Also jet-quenching has been investigated consistently
within a full dynamical picture of the heavy ion collision. The results
for gluonic jets indicate a slightly too large suppression, but are
encouraging to understand the two major phenomena, strong flow and
jet-quenching, within a unified microscopic treatment of kinetic processes.
In addition, simulations
on the temporal propagation of dissipative shock waves lead
to the observation that an $\eta/s$ ratio larger than 0.2
prevents the development of well-defined shock waves 
on timescales typical for ultrarelativistic heavy-ion collisions.
}

\section{Overview}
The large elliptic flow parameter $v_2$ measured by the experiments at
the Relativistic Heavy Ion Collider (RHIC) \cite{rhicv2} suggests that
in the evolving QCD fireball  a fast local equilibration of quarks
and gluons occurs at a very short time scale $\le 1$ fm/c,
and that the locally thermalized state of
matter created, the quark gluon plasma (QGP), behaves as a nearly perfect
fluid. 
In addition, the phenomenon of jet--quenching has been another important
discovery at RHIC \cite{quench}. 
So far, it has not been possible to relate both
phenomena by a common understanding of the underlying microscopic
processes.

In this talk,
we demonstrate that perturbative QCD (pQCD) collisional and, more
importantly,
radiative interactions can 
explain a fast thermalization of the initially nonthermal gluon system, 
the large collective
effects of QGP created at RHIC and the smallness of the shear viscosity
to entropy ratio in a consistent manner by using a relativistic pQCD-based
on-shell parton cascade Boltzmann approach of multiparton scatterings
(BAMPS) \cite{XG05,XG07,EXG08,Ch08,XG08,XGS08,Xu09}. 
Also, due to the inclusion of  radiative
processes, BAMPS can simultaneously account for the quenching of
high-momentum gluons \cite{FXG08}. 
In addition, the possible propagation of dissipative shock waves 
in the QGP can be studied \cite{Bouras:2009nn}.

\section{Elliptic Flow and Shear Viscosity}
BAMPS is a parton cascade, which solves the Boltzmann equation
and is based on
the stochastic interpretation of the transition rate \cite{XG05},
which ensures full detailed balance for multiple scatterings. 
Gluon interactions included  are elastic and screened 
Rutherford-like pQCD $gg\to gg$ 
scatterings as well as pQCD inspired bremsstrahlung
$gg\leftrightarrow ggg$ of Gunion-Bertsch type \cite{XG05,XG07,FXG08}.
In the default simulations,
the initial gluon distributions are taken in a Glauber geometry 
as an ensemble of minijets with
transverse momenta greater than $1.4$ GeV \cite{XG07}, produced via
semihard nucleon-nucleon collisions.
The minijet initial conditions and the subsequent evolution using
the present prescription of BAMPS for two sets of the coupling
$\alpha_s=0.3$ and $0.6$ give nice agreements to the measured 
transverse energy per rapidity over all rapidities
\cite{XGS08,Xu09}.
Other initial conditions incorporating gluon saturation such like
the color glass condensate are also possible \cite{EXG08,Ch08}.

\begin{figure}[t!]
\label{v2shv}
\begin{center}
    \includegraphics[width=7.4cm]{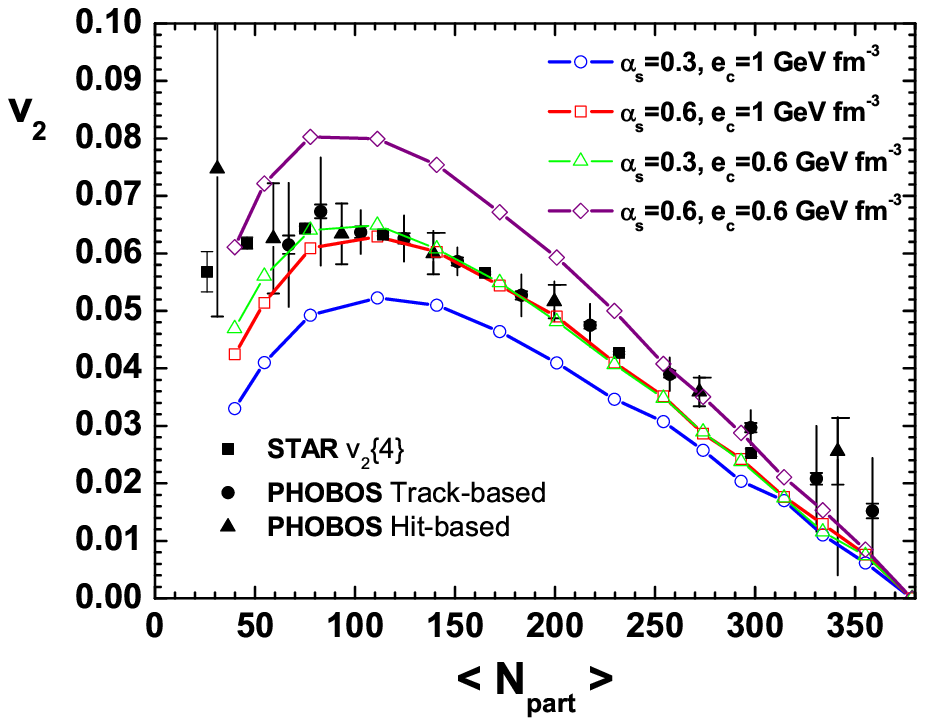}
    \hspace{\fill}
    \includegraphics[width=7.6cm]{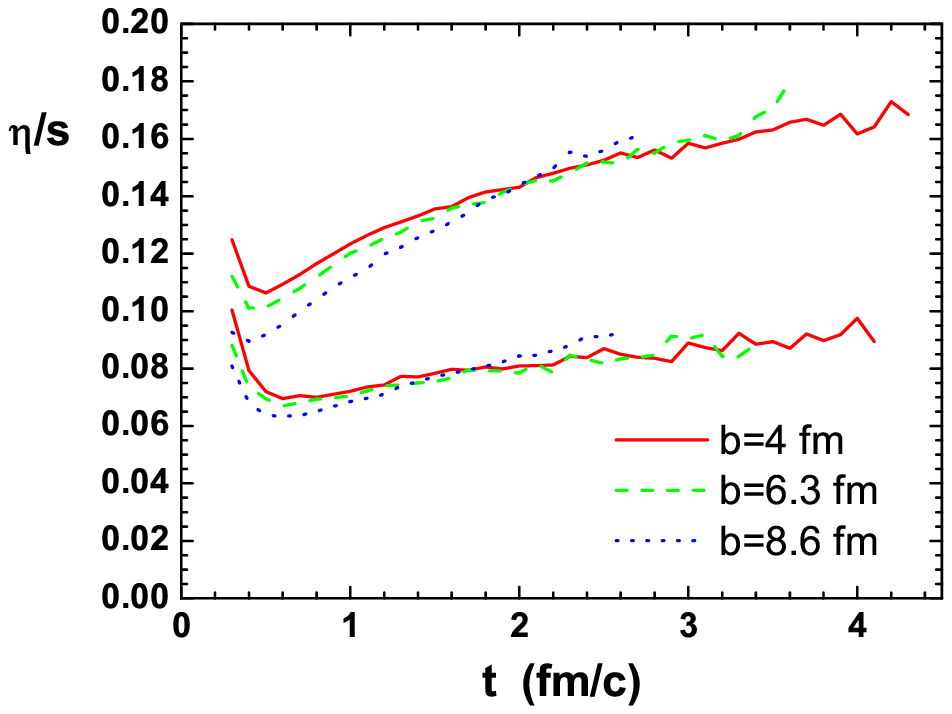}
\caption{(Color online) Left panel: Elliptic flow vs. the number of
participating nucleons for Au+Au collisions
at $\sqrt{s_{NN}}=200$ GeV. The points are STAR  and PHOBOS 
data for charged hadrons within the rapidity $|y| < 0.5$ and $|y| < 1$,
respectively, whereas the curves with symbols are results for gluons within 
$|y| < 1$, obtained from the BAMPS calculations with 
$\alpha_s=0.3$ and $0.6$ and with two freezeout energy densities,
$e_c=0.6$ and $1$ $\rm{GeV fm}^{-3}$.
Right panel: the shear viscosity to entropy density
ratio $\eta/s$ at the central region during the entire expansion. $\eta/s$
values are extracted from the simulations at impact parameter $b=4$,
$6.3$, and $8.6$ fm. The upper band shows the results with $\alpha_s=0.3$
and the lower band the results with $\alpha_s=0.6$.
}
\end{center}
\end{figure}
The elliptic flow $v_2=\langle (p_x^2-p_y^2)/p_T^2 \rangle$ can be
directly calculated from microscopic simulations \cite{XGS08,Xu09}. 
The results are compared with
the experimental data, assuming parton-hadron duality. 
The left panel of Fig. 1 shows the elliptic flow $v_2$ at midrapidity
for various centralities (impact parameters).
Except for the central centrality region either the results with 
$\alpha_s=0.3$ and $e_c=0.6$ ${\rm GeV fm}^{-3}$ (green curves with open
triangles) or those with $\alpha_s=0.6$ and $e_c=1$ ${\rm GeV fm}^{-3}$
(red curves with open squares) agree perfectly with the experimental
data \cite{v2data}.
Within the Navier-Stokes approximation the shear viscosity $\eta$ is
directly related to the so called transport rate \cite{XG08} and
can be calculated dynamically and locally in a full and microscopical
simulation (see the right panel of Fig. 1).
Bremsstrahlung and its back reaction lower the shear viscosity to entropy
density ratio $\eta/s$ significantly by
a factor of $7$, compared with the ratio when only elastic collisions
are considered. According to the extraction by means 
of the full simulation given in Fig. 1, $\eta/s$ is most probably 
lying between $0.2$ for $\alpha_s=0.3$ and $0.08$ for $\alpha_s=0.6$
\cite{EXG08,XG08,XGS08}. The latter value matches
the lower bound of $\eta/s=1/4\pi$ from the AdS/CFT conjecture.
The shear viscosity slightly increases up to $20\%$ when
calculating it by means of Grads momentum method and using the 2nd order
Israel-Stewart framework of dissipative relativistic hydrodynamics \cite{El08}.

\section{Jet-Quenching and Dissipative Shock Waves}
Jet-quenching is generally specified in terms of the nuclear
modification factor $R_{AA}$. We directly compute
$R_{AA}$ by taking the ratio of the final $p_{T}$ spectra to the
initial mini-jet spectra. For this a
suitable weighting and reconstruction scheme of initial jet-like
partons had to be developed 
\cite{FXG08}.  Figure 2 shows the result
for the gluonic contribution to $R_{AA}$, exhibiting a clear
suppression of high--$p_{T}$ gluon jets at a roughly constant level
of $R_{AA}^{\mathrm{gluons}} \approx 0.053$.
\begin{figure}[t!]
\begin{center}
\includegraphics[width=8.8cm]{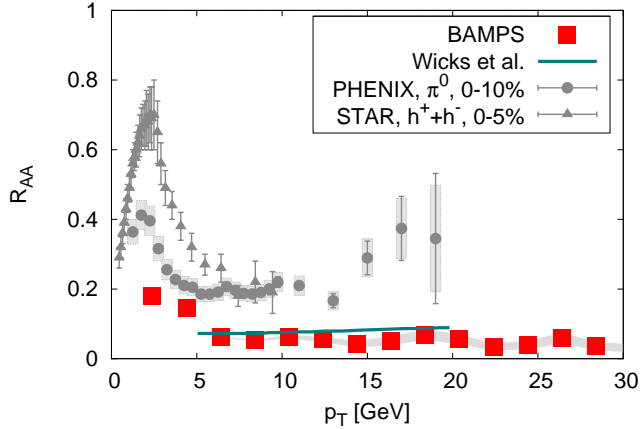}
\caption{(Color online) Gluonic $R_{AA}$ at midrapidity 
($y \,\epsilon\, [-0.5,0.5]$) as extracted from simulations for central 
Au+Au collisions at 200~AGeV. The shaded area indicates the statistical 
error. For direct comparison the result from Wicks et al. 
for the gluonic contribution to $R_{AA}$ and experimental results from 
PHENIX for $\pi^{0}$ and STAR for charged hadrons are shown.
}
\label{fig:RAA}
\end{center}
\vspace{-0.5cm}
\end{figure}

The suppression of gluon jets is approximately a factor $3 \div 4$ stronger
than seen in experimental pion data \cite{rhicjet}. 
An excessive quenching, however, was to be
expected since at present the simulation does not include quarks, which are
bound to lose considerably less energy due to their color factor and dominate the initially produced jets from $p_{T} \approx 20\,\mathrm{GeV}$. 
Indeed, comparing with state of the art results from Wicks et al.
\cite{WHDG07} for the gluonic
contribution to $R_{AA}$ (seen as the line in Fig. \ref{fig:RAA}), which in
their approach together with the quark contribution reproduces the experimental
data, one finds better agreement. The inclusion of light quarks will provide
important means of further verification and will be addressed in a future work.

\begin{figure}[b!]
\includegraphics[width=16.0cm]{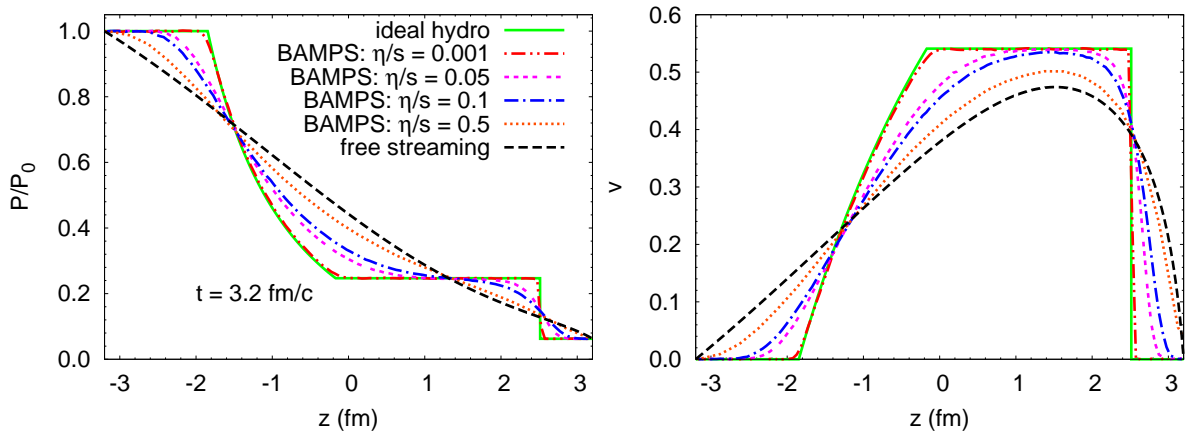}
\caption{(Color online) The pressure (left) and the velocity (right) 
profiles at time $t=3.2$ fm/c.
At $t=0$, the pressure is $P_0=5.43 \ {\rm GeV fm}^{-3}$ for
$z<0$ and $P_4 =0.33 \ {\rm GeV fm}^{-3}$ for $z>0$.
}
\label{fig01}
\end{figure}
In the context of jet-quenching, exciting 
jet-associated particle correlations \cite{Wang:2004kfa} 
have been observed, which indicates the formation of
shock waves in form of Mach cones 
induced by supersonic partons moving through the 
quark-gluon plasma. Shock waves can only form and propagate if
matter behaves like a fluid. At present it is not known whether the
$\eta/s$ value deduced from $v_2$ data is sufficiently small 
to allow for the formation of shock waves.
A step towards this problem is to
consider the relativistic Riemann problem
in viscous gluon matter \cite{Bouras:2009nn}.

In Fig. 3 we show the solution of the Riemann problem for various 
$\eta/s$ values as computed with BAMPS. The results
demonstrate the gradual transition from the 
ideal hydrodynamical limit to free streaming of particles.
A larger $\eta/s$ value results in a finite transition 
layer where the quantities change smoothly rather than 
discontinuously as in the case of a perfect fluid. 
A nonzero viscosity
smears the pressure and velocity profiles and
impedes a clean separation of
the shock front from the rarefaction fan. 
A criterion for a clear separation, and thus the
observability of a shock wave,
is the formation of a shock plateau,
as in the ideal fluid case. The formation of a shock plateau
takes a certain amount of time. From Fig. 3 we infer that, for
$\eta/s>0.1$, a shock plateau has not yet developed at
$t=3.2$ fm/c, whereas for $\eta/s <0.1$, it has already fully
formed. Applying time scaling arguments, 
the formation of shock waves
in gluon matter with $\eta/s > 0.2$ probably takes longer
than the lifetime of the QGP at RHIC \cite{Bouras:2009nn}.

In summary, within the pQCD based parton cascade BAMPS 
early thermalization, the build up of a large 
elliptic flow $v_2$, the smallness of the $\eta/s$ ratio, 
and the quenching of gluonic jets in Au+Au collisions at RHIC
can be understood within one common setup. This is a
committed and large scale undertaking. 
Further analyses on jet quenching,
on particle correlations, on possible Mach cone like behaviour
initiated by a jet, on quark degrees of
freedom including hadronization, on initial conditions and on the
exploration and use of dissipative hydrodynamics are underway to
establish an
even more global picture of heavy ion collisions.

\section*{Acknowledgments}
 This work was supported by
BMBF, DAAD, DFG, GSI and by the 
Helmholtz International Center
for FAIR within the framework of the LOEWE program (Landes-Offensive zur
Entwicklung Wissenschaftlich-\"okonomischer Exzellenz) launched
by the State of Hesse.

\end{document}